\documentclass[fleqn,12pt,twoside]{article}
\usepackage{espcrc1}

\usepackage{graphicx}

\newcommand{\AmS}{{\protect\the\textfont2
  A\kern-.1667em\lower.5ex\hbox{M}\kern-.125emS}}
\def\et{$\it{et~al.}$}
\def\He{$^{5}_{\Lambda}$He}
\def\C{$^{12}_{\Lambda}$C}
\def\B{$^{11}_{\Lambda}$B}
\def\a{$\alpha_{p}^{NM}$}
\def\am{$\alpha^{M}$}
\newcommand{\pik}{($\pi^+$,K$^+$)}

\title{Proton asymmetry in non-mesonic weak decay of light hypernuclei}

\author{T.~Maruta$^a$,~S.~Ajimura$^b$,~K.~Aoki$^c$,~A.~Banu$^d$,~H.~Bhang$^e$,~T.~Fukuda$^f$,~O.~Hashimoto$^g$,
J.~I.~Hwang$^e$,~S.~Kameoka$^g$,~B.~H.~Kang$^e$,~E.~H.~Kim$^e$,~J.~H.~Kim$^e$,~M.~J.~Kim$^e$,
Y.~Miura$^g$,~Y.~Miyake$^a$,~T.~Nagae$^c$,~M.~Nakamura$^a$,~S.~N.~Nakamura$^g$,~H.~Noumi$^c$,
S.~Okada$^h$,~Y.~Okayasu$^g$,~H.~Outa$^i$,~H.~Park$^j$,~P.~K.~Saha$^f$,~Y.~Sato$^c$,~M.~Sekimoto$^c$,
T.~Takahashi$^g$,~H.~Tamura$^g$,~K.~Tanida$^i$,~A.~Toyoda$^c$,~K.~Tsukada$^g$,~T.~Watanabe$^g$,
H.~J.~Yim$^e$\\
\vspace{2mm}
$^a$Department of Physics, University of Tokyo,
Tokyo 113-0033, Japan\\
$^b$Department of Physics, Osaka University,
Osaka 560-0043, Japan\\
$^c$High Energy Accelerator Research Organization,
Tsukuba, Ibaraki 305-0801, Japan \\
$^e$Department of Physics, Seoul National University,
Seoul 151-742, Korea\\
$^d$GSI,
Darmstadt D-64291, Germany\\
$^f$Osaka Electro-Communication University,
Osaka 572-8530, Japan\\
$^g$Physics Department, Tohoku University,
Sendai 980-8578, Japan\\
$^h$Department of Physics, Tokyo Institute of Technology,
Tokyo 152-8551, Japan. \\
$^i$RIKEN, Wako,
Saitama 351-0198, Japan\\
$^i$Korea Research Institute of Standards and Science (KRISS),
Daejeon, 305-600, Korea
}
\begin{document}
%
\maketitle
\begin{abstract}
We have obtained the decay asymmetry parameters in non-mesonic
weak decay of polarized $\Lambda$-hypernuclei by measuring the proton
asymmetry. The polarized $\Lambda$-hypernuclei, {\He}, {\C}, and {\B}, were
produced in high statistics via the \pik\ reaction at 1.05 GeV/$c$ in the forward angles.
Preliminary analysis shows that the decay asymmetry parameters
are very small for these s-shell and p-shell hypernuclei.
\end{abstract}
\section{INTRODUCTION}
The non-mesonic weak decay (NMWD : $\Lambda$N$\rightarrow$NN)
 of $\Lambda$-hypernuclei gives us
a very unique opportunity to study baryon-baryon weak interaction.
There exist several experimental observables; life times(total decay rates), 
branching ratios($\Gamma$($\Lambda$p$\rightarrow$np), 
$\Gamma$($\Lambda$n$\rightarrow$nn)), etc.
The asymmetry parameter of decay proton from the
$\Lambda$p$\rightarrow$np process, {\a}, 
is another important observable to investigate the reaction
mechanism of NMWD, because it
comes from the interference between the parity-conserving and
parity-violating amplitudes.

The decay angular distribution of protons from the NMWD of
polarized hypernuclei, $W$($\theta$), is expressed as
\begin{equation}
  W(\theta)=1+A\cos\theta=1+{\alpha}P_{\Lambda} \cos\theta
  \label{asymdefeq}
\end{equation}
, where $A$ is the asymmetry, $P_{\Lambda}$ denotes the polarization of 
a $\Lambda$-hypernucleus, and
$\theta$ means the emission angle of protons with respect to the polarization axis.

So far, two experiments reported the asymmetry parameters of the
NMWD's of {\C}\cite{e160} and {\He}\cite{e278}.
In KEK-PS E160, a large negative $\alpha_{p}^{NM}$ of -1.3$\pm$0.4 was observed
for p-shell hypernuclei\cite{e160},
while the $\alpha_{p}^{NM}$ of {\He} was measured to be
0.24$\pm$0.22 by KEK-PS E278\cite{e278}.
There is a large discrepancy between two values, which might suggest
the difference of the reaction mechanisms of the NMWD's between
s-shell and p-shell hypernuclei.
However, it would be too early to draw a conclusion in taking account of the
statistical significance of the data and the systematic errors between two
different measurements.
Thus, a new measurement with improved statistics and better systematic
errors had been awaited.
\section{EXPERIMENTS: KEK-PS E462 AND E508}
From year 2000 to 2002, we performed two series of experiments
to measure the NMWD of $^5_{\Lambda}$He (E462) and $^{12}_{\Lambda}$C (E508)
in the  K6 beamline at the KEK 12-GeV PS with the high-resolution and large-acceptance
SKS spectrometer.
The ($\pi^{+}$, K$^{+}$) reactions at 1.05 GeV/$c$ were used
to produce highly-polarized $\Lambda$ hypernuclei with respect to the 
horizontal reaction plane.
Two decay counter systems were installed symmetrically above and below
the target to detect decay protons from NMWD. Each system consisted of a set of drift chambers,
two sets of timing counters and a neutron counter array with six layers of plastic counters.
\begin{figure}[bhtp]
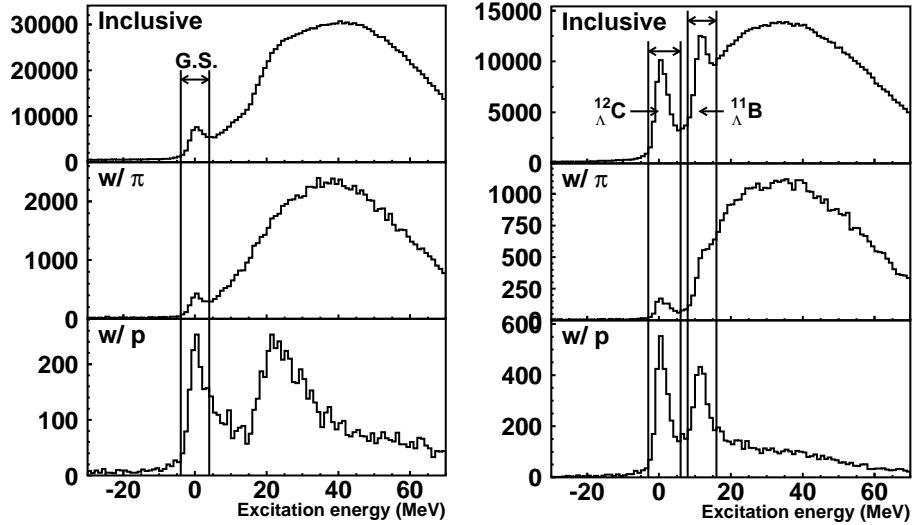

  \begin{center}
    \begin{minipage}{5.9cm}
      \rotatebox{-90}{
	\resizebox{!}{5.8cm}{\includegraphics{mhy-ma_e462_hyp03_2.epsi}}}
      \label{mhy_e462}
    \end{minipage}~
    \begin{minipage}{5.9cm}
      \rotatebox{-90}{
	\resizebox{!}{5.8cm}{\includegraphics{mhy-ma_e508_hyp03_2.epsi}}}
      \label{mhy_e508}
    \end{minipage}
    \caption[fig1]{Excitation energy spectra of {\He} (left) and {\C} (right).
      From top to bottom, inclusive spectra, those with a pion coincidence,
      and with a proton coincidence, respectively.
      \label{mhy}}
  \end{center}
\end{figure}

Figure~\ref{mhy} shows excitation energy spectra of {\He} (left) and {\C} (right)
in three different conditions: from top to bottom, inclusive spectra,
those with a pion coincidence, and with a proton coincidence.
We succeeded to detect 52,000 events of
the {\He} ground state and 67,000 events of the {\C}  ground state,
whose statistics are several times higher than the previous measurements.
The substitutional states of {\C} at around 10 MeV excitation region,
whose selection gate is shown in the figure, are open 
for proton emission, i.e. it goes to {\B}.

\section{PRELIMINARY RESULTS}
The asymmetry, $A$, in Eq. (\ref{asymdefeq}) is obtained from the ratio, $R$,
of numbers of decay particles emitted to the parallel and anti-parallel
directions with respect to that of the hypernuclear polarization.
Owing to the large acceptance of SKS spectrometer,
we can polarize the hypernucleus upward(downward)
by selecting the  ($\pi^+$,K$^+$) scattering direction
on the left(right) at one setup.
Then, the $R$ is expressed as
\begin{equation}
  A = \frac{1-R}{1+R} \quad ,\quad R = 
  \left( \frac
  {N^{+}_{Up}\times N^{-}_{Down}}
  {N^{-}_{Up}\times N^{+}_{Down}}
  \right)^{\frac{1}{2}}\\
  \label{ratioeq}
\end{equation}
, where $N^{+(-)}_{Up(Down)}$ denotes
the yield of the up (down) counter system at the scattering angle
to the left (right). In this expression, systematic errors 
coming from the differences of detection efficiencies and acceptances
between the up and down decay counter systems are canceled out
in the first order. In fact, we found the asymmetries of pions and
protons from the ($\pi^{+}$, p) reaction, in which we don't
expect any asymmetries, were less than 0.3\%.

As shown in Eq. (\ref{asymdefeq}), we need to know the polarization
of a $\Lambda$ hypernucleus to obtain the asymmetry parameter of
proton.
In the case of {\He}, assuming the asymmetry parameter of the
mesonic weak decay, {\am}, is same as that in free space(-0.642$\pm$0.013),
we can measure the polarization of the {\He} from the pion decay
asymmetry.
In the case of p-shell hypernuclei, however, it is hard to estimate
 the polarization from the experimental data, because
 mesonic decay mode is suppressed
due to Pauli blocking, and we don't have a good estimate of {\am}.
Therefore, we needed to estimate the polarization of p-shell hypernuclei 
with a help of a theoretical calculation in Ref.~\cite{itonaga}.
\begin{figure}[hbt]
  \begin{center}
      \resizebox{7.9cm}{!}{\includegraphics{asympara_e462_hyp03_test.epsi}}\\
      \resizebox{7.9cm}{!}{\includegraphics{asympara_e508_hyp03_test.epsi}}
      \caption[fig_e462]{(up) Preliminary result of {\He}
	. The horizontal axis denotes
	polarization, P$_{\Lambda}$, and 
	the vertical one denotes asymmetry, A, for the scattering angles from
	6$^{\circ}$ to 15$^{\circ}$.
	{\a} is obtained from the slope of a linear fitting.
	~(down)  Preliminary result of {\C} and {\B}. each axis is the same
	as that of upper one. Circle points for {\B} and triangle ones for
	{\C}. The final value is derived from a
	weighted average of two hypernculei.
	\label{asympara}}
  \end{center}
\end{figure}

Figure~\ref{asympara} shows a preliminary result
of proton asymmetries in the decay of {\He} measured in different scattering angles, 
6$^{\circ}<|\theta_K|<$15$^{\circ}$,
as a function of the polarization of {\He} obtained from the
mesonic decay.
The slope of the linear fit corresponds to the {\a}, and
we found it to be 0.07$\pm$0.08$^{+0.08}_{-0.00}$.
As a systematic error, we took into account the effect of a small amount of
pion contamination into protons, which tends to reduce the proton asymmetry
to the negative side.
We confirmed the very small positive value of {\a} reported by
Ajimura {\et}\cite{e278} with improved accuracy.
In addition, we found $\alpha_{p}^{NM}$ of p-shell hypernuclei 
obtained from a weighted average of each hypernucleus to be almost zero
(-0.24${\pm}$0.26$^{+0.08}_{-0.00}$) as shown in Fig.~\ref{asympara}.
Note that the present result is twice more accurate than the previous result.
The {\a} (p-shell) of our result is rather consistent with that of {\He}.
It suggests that the decay mechanism of p-shell hypernuclei
is similar to that of s-shell hypernuclei.
\section{SUMMARY}
We have measured the decay asymmetry parameters of proton in the NMWD's of
the polarized hypernuclei, {\He}, {\C}, and {\B}. For the s-shell hypernucleus,
{\He}, it was found to be  0.07$\pm$0.08$^{+0.08}_{-0.00}$, which is 
consistent to the previous value of 0.24$\pm$0.22 \cite{e278} with smaller
errors. For the p-shell hypernuclei, {\C} and {\B}, we found
a similarly small value of -0.24$\pm$0.26$^{+0.08}_{-0.00}$, which largely
deviates from the old value of -1.3$\pm$0.4 \cite{e160}. The new values
are different from those, -0.6$\sim$-0.7, obtained in the recent theoretical calculations 
 \cite{theo1,theo2}. We might need to reconsider the reaction mechanism of the
 NMWD.

%
%

\end{document}